# Noise Spectroscopy and Electrical Transport in $NbO_2$ Memristors with Dual Resistive Switching

*Nitin Kumar, Jong E. Han, Karsten Beckmann, Nathaniel Cady, G. Sambandamurthy[*]*

Nitin Kumar, Jong E. Han, and G. Sambandamurthy
Department of Physics, University at Buffalo-SUNY, Buffalo, NY 14260 US
*E-mail: sg82@buffalo.edu

Karsten Beckmann, and Nathaniel Cady
Department of Nanoscale Science and Engineering, University at Albany, Albany, NY 12202 US

Karsten Beckmann
NY CREATES, Albany, NY 12202 US

Negative differential resistance (NDR) behavior observed in several transition metal oxides is crucial for developing next-generation memory devices and neuromorphic computing systems. $NbO_2$-based memristors exhibit two regions of NDR at room temperature, making them promising candidates for such applications. Despite this potential, the physical mechanisms behind the onset and the ability to engineer these NDR regions remain unclear, hindering further development of these devices for applications. This study employed electrical transport and ultra-low frequency noise spectroscopy measurements to investigate two distinct NDR phenomena in nanoscale thin films of $NbO_2$. By analyzing the residual current fluctuations as a function of time, we find spatially inhomogeneous and non-linear conduction near NDR-1 and a two-state switching near NDR-2, leading to an insulator-to-metal (IMT) transition. The power spectral density of the residual fluctuations exhibits significantly elevated noise magnitudes around both NDR regions, providing insights into physical mechanisms and device size scaling for electronic applications. A simple theoretical model, based on the dimerization of correlated insulators, offers a comprehensive explanation of observed transport and noise behaviors near NDRs, affirming the presence of non-linear conduction followed by an IMT connecting macroscopic device response to transport signatures at atomic level.

1. Introduction

Metal oxide-based memristors exhibiting orders-of-magnitude resistance change through an insulator-to-metal transition (IMT) are excellent candidates for resistive memories and as scalable elements for neuromorphic computing.[1-3] $NbO_2$, $VO_2$, $TaO_x$, and $HfO_2$ have proven to be promising materials as energy-efficient, versatile, and cost-effective resistive memories.[4-9] $NbO_2$ is extensively used as artificial synapses and neurons in neuromorphic computing, probabilistic computing, and as spike generators.[10-12] Despite its broad range of applications, $NbO_2$ remains underexplored for its potential use in multi-level memory applications.

$NbO_2$ undergoes a temperature-driven IMT at ~ 1081 K,[13-14] concomitant with a structural transition from a body-centered tetragonal (BCT) distorted rutile structure (I41/a) to a rutile structure (P42/mnm).[15-16] Two significant mechanisms thought to play key roles in this IMT are Nb-Nb dimer pairing and the softening of vibrational modes.[16-17] For practical applications such as resistive memories, however, the IMT in $NbO_2$ can be triggered by an applied voltage, a phenomenon that has been extensively studied in the context of resistive switching.[1-3, 6] The potential of $NbO_2$ to exhibit multiple resistive switching states, crucial for multi-level memory applications,[18] remains largely unexplored.

Application of a sufficiently large voltage induces a threshold resistive switching behavior in $NbO_2$, which manifests as negative differential resistance (NDR) in current-controlled measurements. In $NbO_2$, investigations have highlighted instances of double NDR phenomena, with NDR-1 attributed to thermally activated Poole-Frankel conduction and NDR-2 linked to the Mott IMT.[19] $NbO_2$, however, does not exhibit strong electronic correlations like its sister compound, $VO_2$;[20] hence, IMT in $NbO_2$ is considered a second-order Peierls transition.[21-23]

Various models, including the electric field-triggered thermal runaway model,[24] trap-assisted Poole-Frankel model,[25] and thermally activated Poole-Frankel model,[19] have been proposed to elucidate NDR in transition metal oxides. Also, many models exclusively address NDR-1, while others propose distinct physical mechanisms for both the NDR-1 and NDR-2.[19] However, these studies often fall short of concluding the type of transition during NDR-2 and establishing a comprehensive link between the conduction mechanisms and device-level observations, which is crucial for effective utilization of such devices.

Highly sensitive, residual fluctuations in electrical measurements (current/voltage) and their analyses in time and frequency domains (known as noise spectroscopy[9, 26-27]) have proven to be valuable assets in elucidating microscopic conduction processes[28-31] in electronic materials as well as for performance assessment of memory devices.[32-33] The presence of a $1/f^\alpha$ -like power spectral density (PSD) has been frequently observed in various resistive switching memories.[32-35] This $1/f^\alpha$ characteristic of the PSD has proved to be particularly sensitive to conduction mechanisms, fabrication processes, and device architecture, thus serving as a valuable tool for establishing crucial connections between mechanisms, material properties, device fabrication and architecture.

In this article, we present results from electrical transport and noise spectroscopy measurements to understand the type of transition and physical mechanisms behind both of the NDR regions in $NbO_2$ thin films. We find that NDR-1 results from spatial inhomogeneous conduction as evidenced by random telegraphic noise in the electrical signal and NDR-2, accompanying the insulator-to-metal transition, is associated with a second-order Peierls transition. Furthermore, a dimer model based on mean-field theory reproduces the experimentally observed non-linear transport characteristics as well as noise spectroscopy results thereby substantiating the experimental findings. These findings for underlying physics of NDR-1 and NDR-2 help to understand the double switching in voltage driven mode which is essential for multi-level memory applications. Finally, the multistate capability is explored, with stable states confirmed by noise spectroscopy. Multiple states increase storage capacity, allowing more data per cell and enhancing area efficiency.[18]

## 2. Results and Discussions

Current-voltage characteristics (IV) of the devices (inset of **Figure 1a** shows the circuit) bring out the interesting transport features in $NbO_2$: As the voltage is ramped up from V = 0 V (voltage-controlled, IV, Figure 1a), non-linear conduction sets in due to Joule heating leading to an inhomogeneous state with electrical domains of varied resistance values.[19, 36] The subsequent proliferation of domains establishes numerous conductive filamentary paths, and multiple filaments can coalesce together to form a single filament[19, 37-39] and manifests as a sudden, significant change in current through the device, marked as $V_{th}$. Previous studies of $NbO_2$ have reported similar filament formation with dimensions ranging from 100 nm to 5 μm.[19, 38-39] The IV characteristics show hysteretic behavior, as the threshold voltage while sweeping the voltage up ($V_{th}$) is distinctly different from the hold voltage ($V_h$) while ramping down, similar to earlier

observation.[19, 38-41]

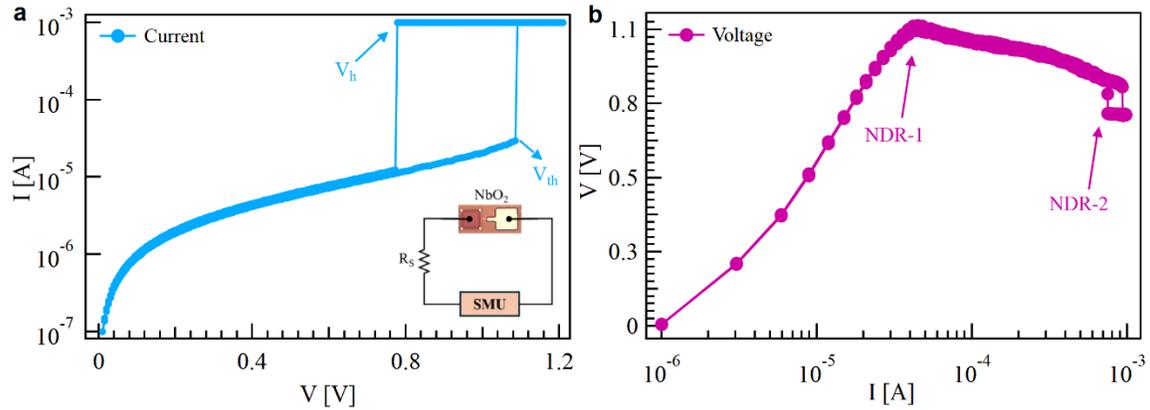

**Figure 1** (a) Current as a function of voltage (IV) from a single NbO$_2$ device, the current was capped at 1.0 mA to protect the device from permanent damage. The inset displays the schematic of the circuit diagram with a microscopic image of NbO$_2$ device. (b) Voltage as a function of current (VI): double NDR was observed in the VI measurement, while threshold switching was observed in the IV measurement. Both measurements are done at room temperature with R$_S$ = 0 Ω.

In the current-driven case (VI, Figure 1b), the current is controlled and hence Joule heating is comparatively lower resulting in a significantly different device response. A negative differential resistance (NDR-1) region sets in first followed by a box-like hysteretic region (NDR-2) at a much higher current value, consistent with earlier observations of two NDRs in NbO$_2$ devices from other reports. [39-40, 42]. The device behavior before the onset of NDR-1 or below V$_{th}$ is likely the same. As the current is increased above NDR-1 (Figure 1b), the gradual increase in power dissipation leads to Joule heating, leading to a temperature increase of several hundred Kelvin in the device. In-situ temperature-dependent X-ray absorption fine structure (XAFS) was used to quantify and correlate the Nb-Nb dimer length as a function of the temperature. This causes orbital overlapping gradually reducing the band gap and eventually at NDR-2 resulting in a structural phase transition closing the bandgap at which point NbO$_2$ becomes metallic.[43]

It is important to understand the subtle differences between the IV and VI characteristics presented in Figure 1. In the IV trace (Figure 1a), only one abrupt jump in current, caused by thermal runway, is observed due to the uncontrollable rise in temperature due to positive feedback from Joule heating. However, in VI (Figure 1b), the current is controlled and hence the Joule heating is limited, and thermal runaway is avoided resulting in the observation of two NDR regions. However, in IV measurements (Figure 1a), a scheme to limit the current in the circuit can provide opportunities to control resistive switching. One possible way to limit the power dissipated in the device in IV (Figure 1b) measurements is through adding a series resistor (inset Figure 1a) to the circuit. Interestingly, the choice of a series resistor can enable control of one vs. two jumps in current, paving the way to utilize the device for storing 2 bits of memory, compared to just 1 bit in devices with a single switching event.[18] Results on how the resistive switching can be precisely controlled are presented in the supporting information (Figure S1), which paves the way for utilizing $NbO_2$ in multistate memory.

Having presented the macroscopic IV and VI characteristics, the focus of this work turned to understanding the microscopic transport behavior in $NbO_2$ devices. Residual electrical fluctuations (noise) and their time and frequency dependence in electronic devices have emerged as important characteristics to obtain valuable insights into the effects of progressive downscaling as well as to delineate and comprehend fundamental resistive switching.[44] Thus, current/voltage noise spectroscopy measurements were carried out in $NbO_2$ thin films under various external parameters with the aim to understand the conduction mechanism in the NDR and non-linear regions in the IV characteristics (Figure 1). Further information about noise measurement and analysis is provided in the supporting information section 2.

As the voltage is ramped up from V = 0 to a value of V = 0.5 V (well below the threshold voltage $V_1$), the residual current fluctuations (**Figure 2**a (i)) result in a relatively lower magnitude of power spectral density (PSD) (**Figure 2**b). These fluctuations follow a Gaussian-like distribution in probability distribution function (PDF), as shown in **Figure 3** a (i). However, as the voltage increases to V = 1.2 V, closer to $V_1$, additional features emerge in the time trace (**Figure 2**a (ii)). The distribution of residual current fluctuations deviates from Gaussianity in PDF(**Figure 3** a (ii)), and the signal resembles random telegraphic noise (RTN), commonly observed in systems near phase transitions.[45-46] This behavior is accompanied by a significant increase in the magnitude of the current fluctuations, leading to an approximately four-order increase in PSD (**Figure 2**b at V = 1.2 V). At V = 1.4 V, the fluctuations return to a Gaussian like distribution PDF (**Figure 3** a (iii)), with a corresponding decrease in PSD (**Figure 2**b), indicating a more uniform conduction behavior in the device. As the bias voltage reaches $V_2$ (1.5 V), anomalous random telegraphic noise (aRTN)[47] is observed in the current fluctuations (**Figure 2**b (iv)), characterized by a bimodal distribution in PDF (**Figure 3** a (iv)) and an increase of up to two orders of magnitude in PSD compared to V = 0.5 V. For V > $V_2$, uniform fluctuations with a Gaussian distribution in

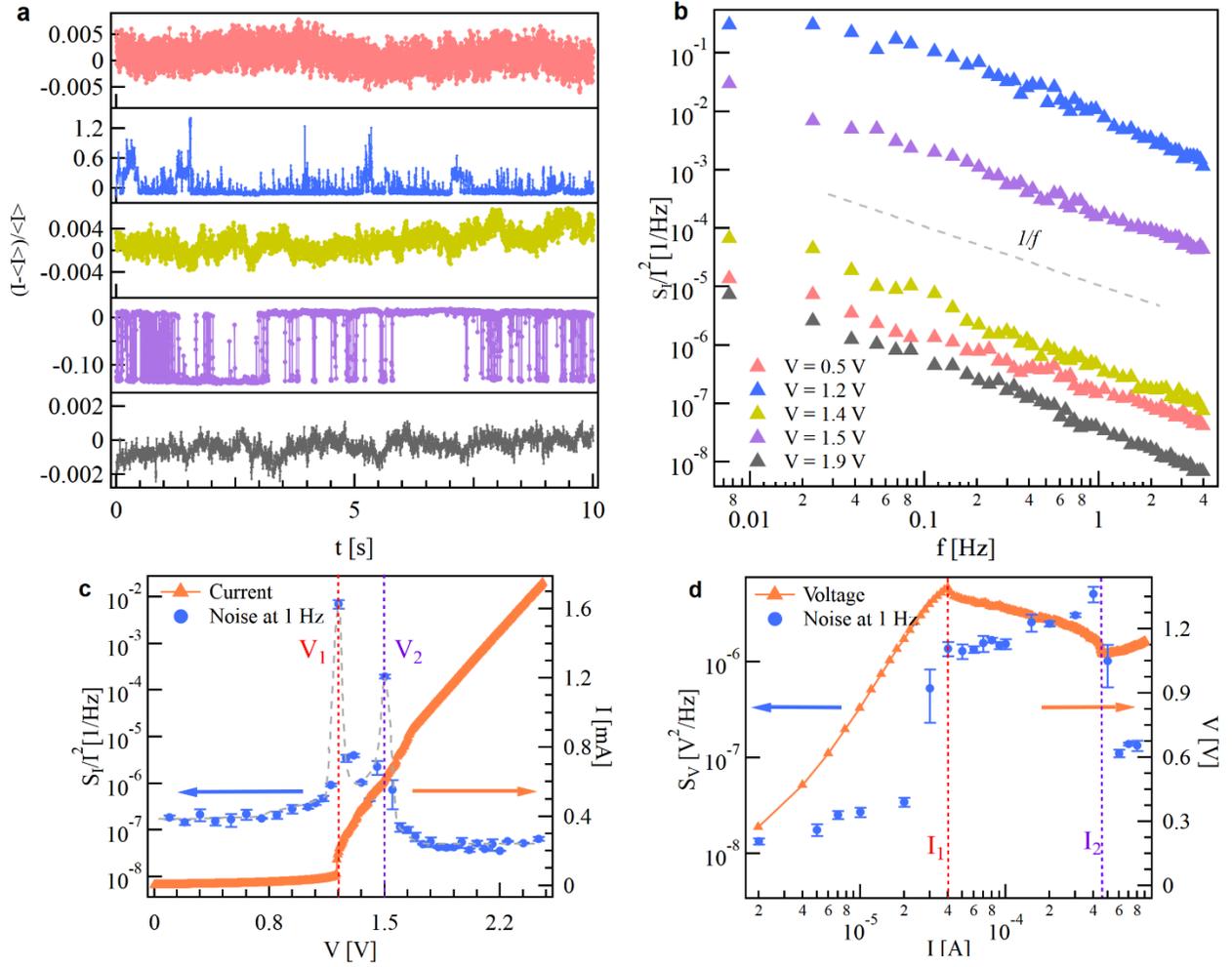

**Figure 2** (a) (i-v) Measured time series of current fluctuations around the mean. Panels (i), (iii), and (v) show fluctuations indicating uniform conduction, while panels (ii) and (iv) display random telegraphic noise, representing the dynamic state of charge carrier transport. (b) Noise power spectral density (PSD) calculated from the time series using Fast Fourier Transformation (FFT) and digital signal processing at five representative voltages. (c) PSD magnitude at 1 Hz as a function of bias voltage shown alongside IV characteristic (right axis) measured with a 1.2 kW series resistor. Vertical lines are guides to indicate the abrupt change in PSD magnitude around the transition point. (d) VI and PSD magnitude at 1 Hz as a function of bias current showing higher noise magnitude between the two NDRs.

reappear (Figure 3 a (v)), along with lower PSD values (Figure 2b), consistent with a stable metallic state. The PSD follows a $1/f^{\alpha}$ trend.

The dependence of the normalized power spectral density ($S_I/I^2$) at a representative frequency of 1 Hz is shown in Figure 2c: the PSD trend observed can be divided into three distinct regions. In the first region ($V \ll V_1$), the PSD remains relatively constant and low in magnitude.

As V approaches $V_1$, a gradual increase in PSD is observed, which can be attributed to spatial inhomogeneities caused by the coexistence of high- and low-conductivity regions, or domains.[38-39, 48] In the second region ($V_1 < V < V_2$), the slope of the IV characteristic becomes steeper, leading to rapid Joule heating. This heating is hypothesized to cause the domains to grow and eventually merge to form a larger domain, surrounded by smaller ones.[38-39] The behavior of PSD in this region will be discussed in more detail later in the manuscript. In the third region ($V > V_2$), the system transitions to a more uniform conducting state akin to metallic systems with insignificant Joule heating, resulting in consistently low PSD values (Figure 2c).

Current-driven noise measurements were performed, allowing for more precise control over the current, to better understand the region between two resistive switching with controlled Joule heating and avoiding thermal runaway scenarios. The magnitude of the power spectral density ($S_V$) at 1 Hz as a function of the source current is shown in **Figure *2*d**. Like the voltage-controlled noise measurements, the conduction mechanism is divided into three regions. The first and third regions away from the transition regions display the same characteristics as seen in the voltage-controlled measurements. In the middle region, the PSD magnitude increases gradually with current, indicating subtle changes in the charge dynamics at the atomic level. Within $NbO_2$, the gradual increase in Nb-Nb dimer length as the temperature increases affects the overlap of Nb orbitals, gradually closing the band gap.[43] This allows more charge carriers to transition from the valence to the conduction band, resulting in an increased PSD magnitude. As the current increases, the device temperature rises due to Joule heating, further closing the band gap through the increased Nb-Nb dimer length, driving the system from its low-temperature insulating phase (body-centered tetragonal) to a high-temperature metallic phase (rutile). This process continues until the transition from the insulating to the metallic phase. The IMT in $NbO_2$ is second order Peierls-dominant, as evidenced by the increase in noise magnitude with rising current, due to the instability of the Nb atomic chain.[23] After this transition, the system enters a homogeneous metallic phase, resulting in the drop in PSD magnitude that remains constant with further increases in current, indicating no further changes in the conduction mechanism. Nb-Nb dimerization has been identified as the key trigger for the onset of the pristine metallic phase, which can be induced through various means such as thermal excitation[43], optical excitation[49], and electrical excitation (this work).

While the noise data in the frequency domain enabled better understanding of the microscopic transport behavior in NbO$_2$ across the different regions, further analysis of the noise data in the time domain, through time-lag plots (TLP), can provide insights into the stable resistive states and transitions among them over the entire voltage range.[50-51] TLPs are simple visualizations of the autocorrelation maps of the one-dimensional time traces (Figure 3b). A third, additional exercise in analyzing noise data is by calculating the probability distribution function (histograms) of the time traces and the results are presented in the **Figure 3** a

The TLPs in **Figure 3** a  are constructed by plotting the value of current at any given time against the value at the previous data point (known as lag order 1). The strength of the autocorrelation is assessed by varying the lag factor. For this dataset, the same pattern is observed consistently as the lag order is increased from 1 to 300 indicating a strongly autocorrelated system.[52] At V= 0.5 V, as seen in **Figure *2*** a (i), the time trace shows random fluctuations around the mean value and the corresponding TLP constructed shows uniform distribution along the diagonal, representative of a single stable state (**Figure 3** a  (i)). Time signals of pink or white noise will display similar characteristics. [53] At these low voltage values, NbO$_2$ has a well-defined band gap (**Figure 3** a (i)) and the physical picture of the device can be represented by a uniform distribution of temperature (**Figure 3** a (i)). As voltage is further increased and approaches V$_1$, the formation of domains due to Joule heating lead to an inhomogeneous system resulting in fluctuations in the band as depicted in **Figure 3** a (ii). It is important to note that in addition to the diagonal distribution in the TLP (**Figure 3** a (ii)), several off-diagonal clusters appear revealing multiple intermediate states that are being formed. Similar RTN in electrical signals have been observed in several systems due to filamentary conduction.[32, 34, 54] This behavior is attributed to inhomogeneous conduction and domains (as depicted in Figure 3d (ii) due to Joule heating, which results in *k*-space fluctuations, as shown in **Figure 3** a  (ii). Following the transition, at V = 1.4 V, a uniform distribution of current fluctuation re-emerges, which could be seen in TLP as uniform distribution along the diagonal (**Figure 3** a (iii)), affirming the uniformity in conduction in real space (**Figure 3** a (iii)) within the device but with a lower and more stable band gap (**Figure 3** a (iii)). At V = V$_2$, the formation of two clusters at diagonal refers to the two-state fluctuations, and transition states can be represented on the band diagram as fluctuation of conduction and valance band above and below the Fermi level, respectively, as shown in **Figure 3** a (iv). Physically, the device temperature increases because of Joule heating and other domain features remain similar (**Figure 3** a (iv)). Any further increase in the applied voltage results in reduced fluctuations,

Gaussian distribution, and a single diagonal cloud in TLP (**Figure 3** a (v)) due to the overlapping

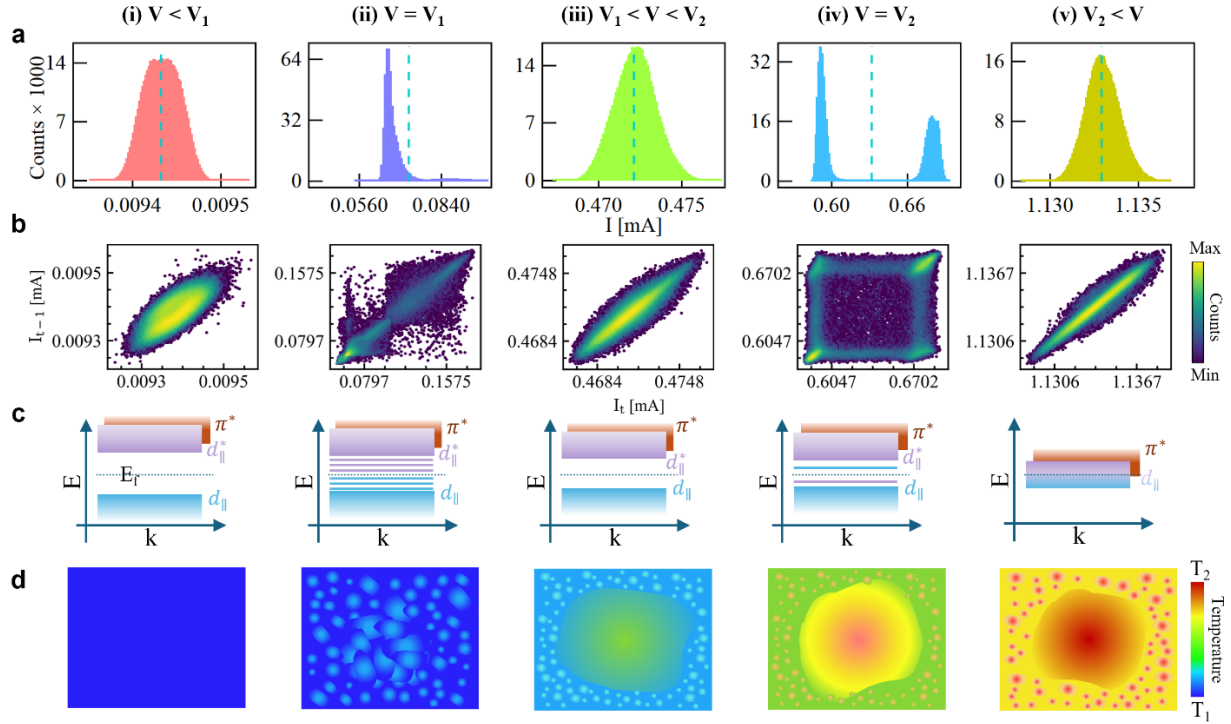

**Figure 3** a (i) Current distribution at V = 0.5 V showing a Gaussian peak centered around 9.40 µA close to the mean value of I. (ii) Current distribution at a voltage of V = 1.2 V, which deviates from Gaussianity, exhibiting a skewed peak around 72 µA and the mean value slightly shifted from the peak. (iii) Current distribution at V = 1.4 V with a Gaussian peak centered around 472 µA close to the mean value of I. (iv) Bimodal current distribution at V = 1.5 V, with non-Gaussian peaks around 600 µA and 690 µA causing aRTS (v) Current distribution at a voltage of V = 1.9 V with a Gaussian peak centered around 1.133 mA close to the mean value of I. The dashed vertical lines indicate the mean current. (b) Time lag plots from noise signals at representative voltages: (i, iii, v) display pink noise-like characteristics generated from uniform conduction, (ii) is from normal random telegraphic noise, (iv) is from an anomalous random telegraphic noise. (c) Schematic representation of the band structure in $NbO_2$ under various voltage conditions: At (i), (iii), and (v), multiple solutions are not feasible, signifying stable conduction. In contrast, at (ii) (NDR-1), fluctuations lead to the existence of multiple insulating states (indicated by dashed lines). At (iv) (NDR-2), fluctuations cause transitions between insulating and metallic states, resulting in a discontinuous shift in the order parameter. At (v), the system stabilizes with overlapping bands at higher voltage. (d) Domain configurations in $NbO_2$ at different bias levels: (i) At $V<V_1$, the system exhibits homogeneous conduction. (ii) As the voltage nears the threshold, multiple domains emerge. (iii) Further voltage increase causes the system to revert to a uniform state dominated by a single large domain. (iv) Near the second resistive switching (NDR-2), two distinct states are possible, as shown by the emergence of separate regions. (v) At high voltage, the system returns to uniformity with increased temperature. The color gradient represents temperature variations from low temperature $T_1$ (blue) to high temperature $T_2$ (red).

of the conduction and valance band resulting in a metallic state in the filament (shown in **Figure**

3 a (v)). The collapse of the band occurs due to the merging of $d_\pi$ and $d_\pi^*$ bands as shown in **Figure 3 a (v)**.[23, 41]

To confirm that the first transition is due to spatial inhomogeneity developing in NbO$_2$ and the second arises from band collapse associated with Nb-Nb dimerization, the resistive switching behavior of NbO$_2$ was further studied with a dimerization model of correlated insulators. The non-equilibrium transport behavior as the device was ramped up from V=0, traversing the two NDR regions (**Figure** ) was self-consistently computed, under a uniform electric field with dissipative thermostats through the Keldysh Green's function theory in the steady-state non-equilibrium limit.[55-58] Motivated by the hysteretic NDR-2 with the collapse of the insulating gap we investigate whether both NDRs can be explained by the same dimerization mechanism. The electronic nature of the NDR states is studied in terms of a spatially uniform mean-field model. The theoretically computed IV (Figure ) does not show any discontinuity at first NDR stage and suggests that the insulating state evolves into a non-linear state possibly leading to thermodynamic inhomogeneity in the sample. On the other hand, NDR-2, the hysteretic transition strongly indicates the eventual destruction of the order and a collapse of the insulating gap in the system, which aligns well with experimental observations. The model consists of two bands[55] ($\alpha$ = 1, 2) on a tight-binding lattice with the self-consistent dimerization order parameter $\Delta$ between the bands controlled by the interacting parameter $U$ in the lattice Hamiltonian with the electron creation/annihilation operators $\left(c_{\alpha i}^\dagger/c_{\alpha i}\right)$ of the band $\alpha$ at the site $i$ with position $x_i$, the tight-binding parameter $t$, the chemical potential $\mu$ the chemical potential $\mu$, and the electric field $E$. $\langle ij \rangle$ denotes summation over nearest neighbors Equation 1. The order parameter $\Delta$ hybridizes the bands and opens an insulating gap and the dissipation to fermion reservoirs[56, 58] is exactly accounted for as a self-energy to the

$$H_{latt} = \sum_\alpha \left[ -t \sum_{\langle ij \rangle} (-1)^\alpha \left(c_{\alpha i}^\dagger c_{\alpha j} + c_{\alpha j}^\dagger c_{\alpha i}\right) + \sum_i \left[(-1)^\alpha (2t - \mu) - E x_i\right] c_{\alpha i}^\dagger c_{\alpha i} \right] \\ + \sum_i \left[ \Delta \left(c_{1i}^\dagger c_{2i} + c_{2i}^\dagger c_{1i}\right) + \frac{\Delta^2}{2U} \right] \quad (1)$$

electron Green's function. To investigate the material properties at multiple stages in the current-controlled case, same circuit geometry[57] as in the experiment is adopted, where an external resistance R$_{ext}$ is connected to the device with a total bias V$_{total}$, and the electric field E in Equation (*1* is determined self-consistently. As the voltage is ramped up (Figure 4a), the current shows a

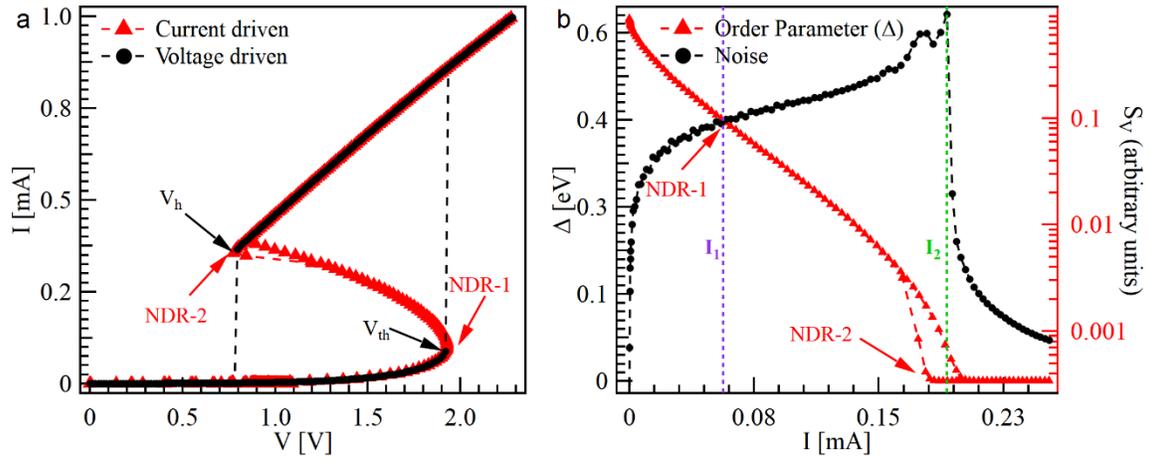

**Figure 4** (a) Shows the relationship between the calculated current vs the device voltage (black curve) and Calculated voltage versus the device current (red curve). (b) Shows the order parameter Δ as a function of total bias (red curve) and Fluctuation in voltage as a function of applied current (black curve). The voltage fluctuation $S_V$ is evaluated via the curvature of the free-energy minimum. See supporting information Section 3 for details.

non-linear behavior that eventually transitions to NDR-1 or $V_{th}$. The agreement of the IV with a mean-field theory and its smooth behavior at the NDR-1 (Figure ) indicate that the electronic states are well-defined with a stable bandgap. The source of the measured large fluctuations is from macroscopic domain fluctuations due to thermodynamic degeneracy at NDR-1 at which multiple solutions are permitted. With further increase in voltage, calculations show a discontinuous transition resulting in a hysteretic IV loop with two states (marked as NDR-2) which can be attributed to the IMT in $NbO_2$. Despite the lack of spatial dynamics in the model, the demonstration of NDR-2 is consistent with experimental observations. To confirm the IMT, order parameter is calculated which represents the band gap. The computed order parameter (Δ) plotted against the total voltage (Figure ). Within the model, an increase in current enhances charge fluctuations, which softens the order parameter Δ and the insulating gap. This rapidly reduces the system's resistance $R(I)$ and the device voltage $IR(I)$, leading to the NDR-1 behavior. This NDR region from $I_1$ persists until current reaches $I_2$ where the order parameter eventually destroyed in a discontinuous transition, resulting IMT marked as NDR-2 (Figure 4b). We also demonstrate the enhanced noise (black curve in Figure ) in the region between the $I_1$ and $I_2$. In the region between $I_1$ and $I_2$, the free-energy has shallow minimum before the collapse of the gap, leading to strong current noise close to NDR-2. Such states, however, have a shallow free-energy minimum which leads to strong fluctuations in current (as plotted in Figure ) in qualitative agreement with Figure

## 3. Conclusions

Non-linear conduction mechanisms and resistive switching were investigated in NbO$_2$ thin films using electrical transport measurements and noise spectroscopy measurements. First resistive switching is attributed to inhomogeneous conduction in the low-voltage region, likely due to domain formation as evidenced by random telegraphic noise in noise signature. In between two resistive switching states, Joule heating plays a major role in the conduction mechanism. At the second resistive switching level, the band gap closes due to an insulator-to-metal transition, resulting in a two-state fluctuation. Additionally, current-driven measurements confirm the Peierls-dominant nature of the insulator-to-metal transition in NbO$_2$ due to the role of Nb-Nb dimerization, which leads to band gap closure in the transition. These results, therefore, extend the insight into the atomic mechanisms of NDR-1 and NDR-2 phenomena and provide pathways to applications of NbO$_2$ based devices in multi-level memory and neuromorphic computing applications.

## 4. Experimental Section

*Device Fabrication*: Electrically measurable nanoscale NbO$_2$ devices were fabricated using the NY CREATES / UAlbany memory test vehicle (MTV) with device diameter of ~120 nm. The MTV provides a W interconnect layer beneath a nanoscale TiN bottom electrode (BE) embedded in Si$_3$N$_4$. The device stack is deposited in a PVD75 tool from Kurt J. Lesker Company via reactive Ar sputtering in a partial $O_2$ atmosphere for $NbO_2$ followed by ex-situ anneal at 750 °C for 5 min and subsequently capped via Ar sputtering with 5 nm Ir and 100nm W. The $NbO_2$ thin film is doped with Ti by using a $Ti_{0.1}Nb_{0.9}$ sputter target which stabilizes the dioxide phase. After the deposition over the nanoscale TiN BE, a single patterning step via contact lithography and a reactive ion etch (RIE) is used to isolate the devices enabling the electrical contact to the W interconnect layer and the W top contact (Figure ).[59-60]

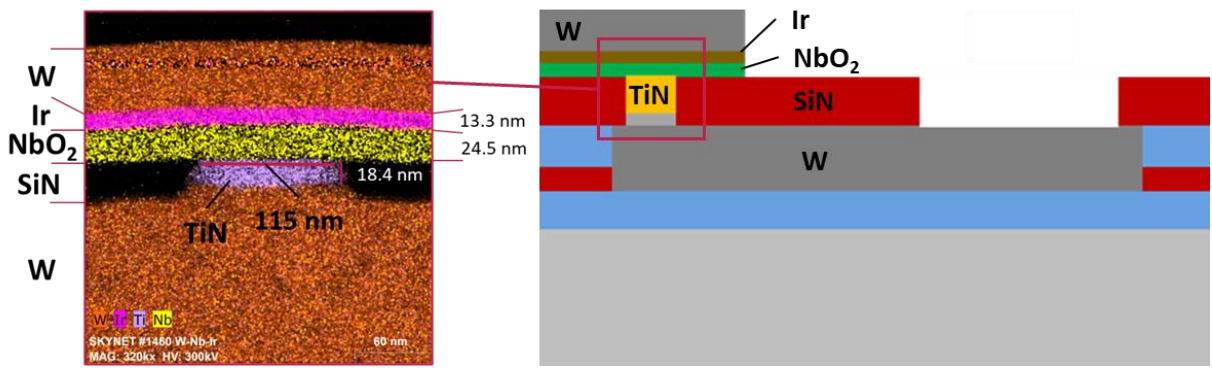

**Figure 5**: Illustration (right) and false-colored EDX map on a TEM micrograph (left) for the nanoscale $NbO_2$ devices integrated onto the NY CREATES / UAlbany Memory Test Vehicle. The $NbO_2$ thin film is embedded between a TiN and Ir bottom and top electrode, respectively.

*Electrical Measurements and Characterization*: Electrical transport measurements were carried out using probe stations Rucker and Kolls Model 260 or SUSS MicroTec. Current-Voltage characteristics were performed using a Keithley 2450 Source Meter Unit (SMU) using Test Script Processor (TSP) commands.

# Supporting Information

1. CONTROLLING RESISTIVE SWITCHING WITH EXTERNAL RESISTORS

A series resistor allows to alter the single threshold switching and to achieve double resistive switching in the current-voltage (IV) characteristics, by ensuring that the total current is limited preventing a thermal runaway scenario. IV measurements with a series resistor of different resistance values are shown in Figure S. A single resistive switching at a threshold voltage was observed while using resistors up to 800 Ω, but two resistive switching events appeared for higher resistance values. The derivative of IV (inset of Figure S) was used to distinguish between single and double resistive switching. Controlling the current in the circuit helps manage thermal runaway and allows active regulation of resistance changes at threshold voltage levels. This approach opens new possibilities for incorporating $NbO_2$ in multi-level RAM devices.

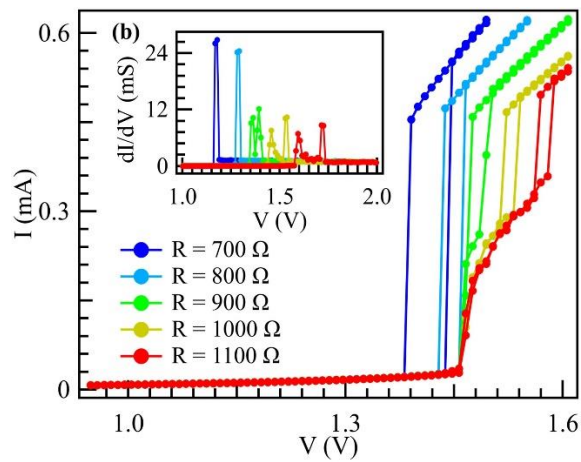

Figure S1 The IV characteristics with a set of series resistors. The inset shows differential conductance as a function of voltage (plots shifted leftward for clarity).

2. NOISE SPECTROSCOPY MEASUREMENTS

Noise spectroscopy measurements contain three main steps: time-dependent data acquisition, signal processing, and analysis. [31, 45] During the first step, the sample is placed on a probe station platform and connected to a Source Measure Unit (SMU) via a series resistor ($R_s$). Voltage/current is sourced and measured using SMU, which measures and records current/voltage as a function of time at a sampling rate of 512 points per second.

In the signal processing step, background noise is filtered out using digital processing techniques. By applying digital filters, unwanted frequency components are removed, preserving the relevant signal information. This step is followed by decimation, which reduces the sampling rate to decrease the data size without losing significant information, improving the efficiency of further analysis. Detrending is then applied to isolate the actual noise signal for accurate analysis, removing trends and drifts unrelated to the sample's inherent characteristics and mean of the fluctuations are subtracted from the signal, resulting in fluctuations centered around the mean..

The final step, analysis, can be done in time domain as well as in frequency domain. The processed signal can be converted from the time domain to the frequency domain using Fast Fourier Transform (FFT). This transformation reveals the signal's frequency components, essential for understanding its spectral characteristics. Power Spectral Density (PSD) helps to quantify how the signal's power is distributed across various frequencies. The time domain signal allows us to understand the temporal characteristics such as telegraphic signal, multi-state switching, etc.

Another useful tool to characterize electronic noise in resistive switching devices is the calculation of probability distribution function of the fluctuations. Gaussian distribution is an indication of a random and uncorrelated process, while deviation from Gaussianity points to underlying correlations. Such a temporal correlation and repetitive structure in noise signal can be understood as illustrated in the time-lag plots (TLP) in main text. Time-lag plots were constructed by plotting the magnitude of the signal at a particular time against its value after a chosen "lag" in time. Time-lag plots are useful to check whether the time series of fluctuations is random or any identifiable structure. For example, clusters of points along the diagonal in the plot correspond to stable states of the system while off-diagonal clusters point to transitions between the stable states.[50-51]

## 3. DIMERIZATION MODEL OF CORRELATED INSULATORS

We interpret the nonequilibrium phase transition based on the steady-state nonequilibrium Green's function (NEGF) theory[56, 58] with the Hamiltonian Eq. (1). Within the mean-field theory, [56] the order parameter $\Delta$ is self-consistently updated in the device electric-field $E$ to a new order

parameter as $\Delta'$

$$\Delta' = U\langle c_{1i}^\dagger c_{2i} + c_{2i}^\dagger c_{1i}\rangle(\Delta, E) \qquad (1)$$

where the right-hand-side is evaluated by the NEGFs as a function of $\Delta$ and $E$. The details of the NEGF theory have been presented elsewhere.[55-58] Similarly, the device electric field $E$ is subject to the condition set by an external resistor $R_{ext}$ as $V_{tot} = R_{ext}I(E,\Delta) + LE$ with the device length $L$ and the current $I(E,\Delta)$. The equation is converted to express the updated device electric field $E$ given by the current density in the device $J(E,\Delta) = I(E,\Delta)/A$, as

$$E' = E_{tot} + \left(\frac{AR_{ext}}{L}\right)J(E,\Delta), \qquad (2)$$

with $E_{tot}$ defined as $V_{tot}/L$. The device current density $J(E,\Delta)$ is also computed directly from the NEGF.[58] In the simulation, we fix the parameter $\rho_{ext} \equiv AR_{ext}/L$ and sweep $E_{tot}$ for self-consistent values for $E$ and $\Delta$ through Eqs. (2, 3).

For the sake of brevity, we rewrite the above equations as

$$\Delta' = f(\Delta, E) \text{ and } E' = g(\Delta, E), \qquad (3)$$

respectively. About a fixed point solution $(\Delta^*, E^*)$, the deviation of the variables $\Delta$ and $E$ can be written as $d\Delta' = f_1 d\Delta + f_2 dE$ and $dE' = g_1 d\Delta + g_2 dE$ with the derivatives at the fixed point $f_1 = \left(\frac{\partial f}{\partial \Delta}\right)^*$, $f_2 = \left(\frac{\partial f}{\partial E}\right)^*$, etc. We intend to reduce the above relations to thermodynamic fluctuations in the order parameter $\Delta$ via an effective free energy, to interpret the measured voltage fluctuations. Therefore, we define an effective free energy $F(\Delta)$ with its derivative given as [56]

$$\frac{dF}{d\Delta} = \Delta - f(\Delta, E(\Delta)), \qquad (4)$$

so that the free-energy minimum is equivalent to the self-consistent mean-field condition. To achieve this, we eliminate $E$ by setting $dE' = dE$ in Eq. (3) with the result, $dE = \left(\frac{g_1}{1-g_2}\right)d\Delta$. Substituting this to Eq. (2), we obtain

$$\frac{d^2F}{d\Delta^2} = 1 - f_1 + \frac{g_1 f_2}{g_2}. \qquad (5)$$

The inverse of the free-energy at a local minimum gives us the statistical fluctuations of the order parameter up to an unspecified constant

$$\langle(\delta\Delta)^2\rangle \propto \left(1 - f_1 + \frac{g_1 f_2}{g_2}\right)^{-1}. \tag{6}$$

Finally, to convert this to the device voltage fluctuations, we multiply the factor $\left(\frac{dE}{d\Delta}\right)^2 = \left(\frac{g_1}{1-g_2}\right)^2$ to obtain

$$\langle(\delta V)^2\rangle \propto \left(\frac{g_1}{1-g_2}\right)^2 \left(1 - f_1 + \frac{g_1 f_2}{g_2}\right)^{-1}. \tag{7}$$

It is important to note that, while this mean-field approach gives us a first-hand description of the transport noise in the device, it assumes Gaussian fluctuations around a well-defined local minimum. The mean-field conditions, Eqs. (2, 3), do not have access to domain fluctuations in the vicinity of instabilities. The calculation shows large fluctuations in the second NDR due to the destabilization of the insulating free-energy minimum. However, it underestimates the strong domain fluctuations at the tight turn of the IV curve near the first NDR.

### Acknowledgements

Electrical transport measurements at the University at Buffalo were supported by the National Science Foundation grant # NSF-MRI 1726303.

NK acknowledges support from the College of Arts and Sciences at the University at Buffalo.

Support for device fabrication efforts at the University at Albany was provided by the Air Force Research Laboratory, award # FA8750-21-1-1019.

### References
[1]   M. D. Pickett, G. Medeiros-Ribeiro, R. S. Williams, *Nature Materials* **2013**, 12, 114.
[2]   S. H. Jo, T. Chang, I. Ebong, B. B. Bhadviya, P. Mazumder, W. Lu, *Nano Letters* **2010**, 10, 1297.
[3]   S. Kumar, R. S. Williams, Z. W. Wang, *Nature* **2020**, 585, 518.
[4]   S. M. Yu, H. Y. Chen, B. Gao, J. F. Kang, H. S. P. Wong, *ACS Nano* **2013**, 7, 2320.
[5]   L. Pellegrino, N. Manca, T. Kanki, H. Tanaka, M. Biasotti, E. Bellingeri, A. S. Siri, D. Marré, *Advanced Materials* **2012**, 24, 2929.

[6]  X. Liu, S. M. Sadaf, M. Son, J. Park, J. Shin, W. Lee, K. Seo, D. Lee, H. Hwang, *IEEE Electron Device Letters* **2012**, 33, 236.
[7]  S. H. Bae, S. Lee, H. Koo, L. Lin, B. H. Jo, C. Park, Z. L. Wang, *Advanced Materials* **2013**, 25, 5098.
[8]  S. Sihn, W. L. Chambers, M. Abedin, K. Beckmann, N. Cady, S. Ganguli, A. K. Roy, *Small* **2024**, DOI: https://doi.org/10.1002/smll.2023105422310542.
[9]  D. M. Fleetwood, *IEEE Transactions on Nuclear Science* **2015**, 62, 1462.
[10] H. Rhee, G. Kim, H. Song, W. Park, D. H. Kim, J. H. In, Y. Lee, K. M. Kim, *Nature Communications* **2023**, 14, 7199.
[11] S. Choi, S. Jang, J. H. Moon, J. C. Kim, H. Y. Jeong, P. Jang, K. J. Lee, G. Wang, *NPG Asia Materials* **2018**, 10, 1097.
[12] X. M. Zhang, Y. Zhuo, Q. Luo, Z. H. Wu, R. Midya, Z. R. Wang, W. H. Song, R. Wang, N. K. Upadhyay, Y. L. Fang, F. Kiani, M. Y. Rao, Y. Yang, Q. F. Xia, Q. Liu, M. Liu, J. J. Yang, *Nature Communications* **2020**, 11, 51
[13] K. Seta, K. Naito, *Journal of Chemical Thermodynamics* **1982**, 14, 921.
[14] R. Pynn, J. D. Axe, *Journal of Physics C-Solid State Physics* **1976**, 9, L199.
[15] R. Pynn, J. D. Axe, R. Thomas, *Physical Review B* **1976**, 13, 2965.
[16] A. K. Cheetham, C. N. R. Rao, *Acta Crystallographica Section B-Structural Science* **1976**, 32, 1579.
[17] S. M. Shapiro, J. D. Axe, G. Shirane, P. M. Raccah, *Solid State Communications* **1974**, 15, 377.
[18] B. Sun, S. Ranjan, G. Zhou, T. Guo, Y. Xia, L. Wei, Y. N. Zhou, Y. A. Wu, *Materials Today Advances* **2021**, 9, 100125.
[19] S. Kumar, Z. W. Wang, N. Davila, N. Kumari, K. J. Norris, X. P. Huang, J. P. Strachan, D. Vine, A. L. D. Kilcoyne, Y. Nishi, R. S. Williams, *Nature Communications* **2017**, 8, 658.
[20] F. Gervais, W. Kress, *Physical Review B* **1985**, 31, 4809.
[21] A. O'Hara, A. A. Demkov, *Physical Review B* **2015**, 91, 094305.
[22] K. Kulmus, S. Gemming, M. Schreiber, D. Pashov, S. Acharya, *Physical Review B* **2021**, 104, 035128.
[23] M. J. Wahila, G. Paez, C. N. Singh, A. Regoutz, S. Sallis, M. J. Zuba, J. Rana, M. B. Tellekamp, J. E. Boschker, T. Markurt, J. E. N. Swallow, L. A. H. Jones, T. D. Veal, W. L. Yang, T. L. Lee, F. Rodolakis, J. T. Sadowski, D. Prendergast, W. C. Lee, W. A. Doolittle, L. F. J. Piper, *Physical Review Materials* **2019**, 3, 074602.
[24] C. Funck, S. Menzel, N. Aslam, H. H. Zhang, A. Hardtdegen, R. Waser, S. Hoffmann-Eifert, *Advanced Electronic Materials* **2016**, 2, 1600169.
[25] G. A. Gibson, S. Musunuru, J. M. Zhang, K. Vandenberghe, J. Lee, C. C. Hsieh, W. Jackson, Y. Jeon, D. Henze, Z. Y. Li, R. S. Williams, *Applied Physics Letters* **2016**, 108, 023505.
[26] C. Ciofi, B. Neri, *Journal of Physics D-Applied Physics* **2000**, 33, R199.
[27] L. K. J. Vandamme, *IEEE Transactions on Electron Devices* **1994**, 41, 2176.
[28] S. Kogan, *Electronic Noise and Fluctuations in Solids*, Cambridge University Press, Cambridge **1996**.
[29] Y. Song, S. I. Lee, J. R. Gaines, *Physical Review B* **1992**, 46, 14.
[30] M. B. Weissman, *Reviews of Modern Physics* **1988**, 60, 537.
[31] A. M. Alsaqqa, S. Singh, S. Middey, M. Kareev, J. Chakhalian, G. Sambandamurthy, *Physical Review B* **2017**, 95, 125132.
[32] J. K. Lee, H. Y. Jeong, I. T. Cho, J. Y. Lee, S. Y. Choi, H. I. Kwon, J. H. Lee, *IEEE Electron Device Letters* **2010**, 31, 603.
[33] S. B. Lee, S. Park, J. S. Lee, S. C. Chae, S. H. Chang, M. H. Jung, Y. Jo, B. Kahng, B. S. Kang, M. J. Lee, T. W. Noh, *Applied Physics Letters* **2009**, 95, 122112.
[34] S. M. Yu, R. Jeyasingh, Y. Wu, H. S. P. Wong, *Physical Review B* **2012**, 85, 045324.
[35] M. S. Lee, J. K. Lee, H. S. Hwang, H. C. Shin, B. G. Park, Y. J. Park, J. H. Lee, *Japanese*

[35] *Journal of Applied Physics* **2011**, 50, 011501.
[36] K. Szot, W. Speier, G. Bihlmayer, R. Waser, *Nature Materials* **2006**, 5, 312.
[37] H. Ahn, K. Kang, Y. Song, W. Lee, J. K. Kim, J. Kim, J. Lee, K. Y. Baek, J. Shin, H. Lim, Y. Kim, J. S. Lee, T. Lee, *Advanced Functional Materials* **2022**, 32, 2107727.
[38] S. Kumar, R. S. Williams, *Nature Communications* **2018**, 9, 2030.
[39] S. K. Nandi, E. Puyoo, S. K. Nath, D. Albertini, N. Baboux, S. K. Das, T. Ratcliff, R. G. Elliman, *ACS Applied Materials & Interfaces* **2022**, 14, 29025.
[40] S. K. Nandi, S. K. Nath, A. E. El-Helou, S. Li, X. J. Liu, P. E. Raad, R. G. Elliman, *Advanced Functional Materials* **2019**, 29, 1906731.
[41] P. Chen, X. M. Zhang, Q. Liu, M. Liu, *Applied Physics A* **2022**, 128, 1113.
[42] S. Kumar, J. P. Strachan, R. S. T. Williams, *Nature* **2017**, 548, 318.
[43] G. J. P. Fajardo, S. A. Howard, E. Evlyukhin, M. J. Wahila, W. R. Mondal, M. Zuba, J. E. Boschker, H. Paik, D. G. Schlom, J. T. Sadowski, S. A. Tenney, B. Reinhart, W. C. Lee, L. F. J. Piper, *Chemistry of Materials* **2021**, 33, 1416.
[44] J. K. Lee, I. T. Cho, H. I. Kwon, C. S. Hwang, C. H. Park, J. H. Lee, *IEEE Electron Device Letters* **2012**, 33, 1063.
[45] S. Ghosh, Z. E. Nataj, F. Kargar, A. A. Balandin, *ACS Applied Materials & Interfaces* **2024**, 16, 20920.
[46] A. Bid, A. Guha, A. K. Raychaudhuri, *Physical Review B* **2003**, 67, 174415.
[47] M. J. Uren, M. J. Kirton, S. Collins, *Physical Review B* **1988**, 37, 8346.
[48] S. Singh, G. Horrocks, P. M. Marley, Z. Z. Shi, S. Banerjee, G. Sambandamurthy, *Physical Review B* **2015**, 92, 155121.
[49] R. Rana, J. M. Klopf, J. Grenzer, H. Schneider, M. Helm, A. Pashkin, *Physical Review B* **2019**, 99, 041102.
[50] T. Nagumo, K. Takeuchi, S. Yokogawa, K. Imai, Y. Hayashi, *2009 IEEE International Electron Devices Meeting (IEDM), Baltimore* **2009**, 709.
[51] F. M. Puglisi, P. Pavan, A. Padovani, L. Larcher, G. Bersuker, *Solid-State Electronics* **2013**, 84, 160.
[52] A. P. Nicolau, K. Dyson, D. Saah, N. Clinton, in *Cloud-Based Remote Sensing with Google Earth Engine: Fundamentals and Applications* (Eds: J. A. Cardille, M. A. Crowley, D. Saah, N. E. Clinton), Springer International Publishing, Cham **2024**, p. 403.
[53] M. Robitaille, H. Yang, L. Wang, B. W. Deng, N. Y. Kim, *Scientific Reports* **2023**, 13, 10403
[54] Y. Song, H. Jeong, S. Chung, G. H. Ahn, T. Y. Kim, J. Jang, D. Yoo, H. Jeong, A. Javey, T. Lee, *Scientific Reports* **2016**, 6, 33967
[55] J. E. Han, C. Aron, X. Chen, I. Mansaray, J. H. Han, K. S. Kim, M. Randle, J. P. Bird, *Nature Communications* **2023**, 14, 2936
[56] J. J. Li, C. Aron, G. Kotliar, J. E. Han, *Physical Review Letters* **2015**, 114, 226403.
[57] J. J. Li, C. Aron, G. Kotliar, J. E. Han, *Nano Letters* **2017**, 17, 2994.
[58] J. E. Han, J. J. Li, C. Aron, G. Kotliar, *Physical Review B* **2018**, 98, 035145.
[59] Z. R. Robinson, K. Beckmann, J. Michels, V. Daviero, E. A. Street, F. Lorenzen, M. C. Sullivan, N. Cady, A. Kozen, M. Currie, (Preprint), 2024, arXiv:2406.13523.
[60] M. C. Sullivan, Z. R. Robinson, K. Beckmann, A. Powell, T. Mburu, K. Pittman, N. Cady, *Journal of Vacuum Science & Technology B* **2022**, 40, 063202.